\documentclass[12pt,twoside]{article}
\usepackage[latin1]{inputenc}
\usepackage{times}
\usepackage{german}
\usepackage{epsfig}
\usepackage{amsfonts}
\usepackage{latexsym}
\usepackage{float}
\usepackage{amsmath}
\usepackage{lscape}
\usepackage{pdflscape}
\usepackage{rotating}
\usepackage{nicefrac} 
\usepackage{graphicx,color}

\relax

\relax

\newcommand{\RA}{\Rightarrow}

\newcommand{\sep}{\quad}

\newcommand{\mc}{\multicolumn}

\newcommand{\bay}{\begin{array}}
\newcommand{\eay}{\end{array}}

\newcommand{\bsl}{\begin{slide}}
\newcommand{\esl}{\end{slide}}

\newcommand{\bfr}{\begin{frame}}
\newcommand{\efr}{\end{frame}}

\newcommand{\bol}{\begin{overlay}}
\newcommand{\eol}{\end{overlay}}

\newcommand{\bqa}{\begin{eqnarray*}}
\newcommand{\eqa}{\end{eqnarray*}}

\newcommand{\bqan}{\begin{eqnarray}}
\newcommand{\eqan}{\end{eqnarray}}

\newcommand{\bqt}{\begin{quote}}
\newcommand{\eqt}{\end{quote}}

\newcommand{\bt}{\begin{tabbing}}
\newcommand{\et}{\end{tabbing}}

\newcommand{\bit}{\begin{itemize}}
\newcommand{\eit}{\end{itemize}}

\newcommand{\bist}{\begin{itemstep}}
\newcommand{\eist}{\end{itemstep}}

\newcommand{\ben}{\begin{enumerate}}
\newcommand{\een}{\end{enumerate}}

\newcommand{\beq}{\begin{equation}}
\newcommand{\eeq}{\end{equation}}

\newcommand{\bdes}{\begin{description}}
\newcommand{\edes}{\end{description}}

\newcommand{\btb}{\begin{tabular}}
\newcommand{\etb}{\end{tabular}}

\newcommand{\bpic}{\begin{picture}}
\newcommand{\epic}{\end{picture}}

\newcommand{\bcen}{\begin{center}}
\newcommand{\ecen}{\end{center}}

\newcommand{\bfg}{\begin{figure}}
\newcommand{\efg}{\end{figure}}

\newcommand{\bmp}{\begin{minipage}}
\newcommand{\emp}{\end{minipage}}

\newcommand{\bgan}{\begin{gather}}
\newcommand{\egan}{\end{gather}}

\newcommand{\bal}{\begin{align*}}
\newcommand{\eal}{\end{align*}}

\newcommand{\baln}{\begin{align}}
\newcommand{\ealn}{\end{align}}

\newcommand{\bala}{\begin{alignat*}}
\newcommand{\eala}{\end{alignat*}}

\newcommand{\balan}{\begin{alignat}}
\newcommand{\ealan}{\end{alignat}}

\newcommand{\bspt}{\begin{split}}
\newcommand{\espt}{\end{split}}

\newcommand{\bvt}{\begin{verbatim}}
\newcommand{\evt}{\end{verbatim}}

\newcommand{\lam}{\lambda}

\newtheorem{definition}{{\sc Definition}\sc}[section]
\newcommand{\bdefi}{\begin{definition}}
\newcommand{\edefi}{\end{definition}}

\newtheorem{appropr}[definition]{{\sc Approximation Procedure}\sc}
\newcommand{\bappr}{\begin{appropr}}
\newcommand{\eappr}{\end{appropr}}

\newtheorem{bedi}[definition]{{\sc Condition}\sc}
\newcommand{\bbd}{\begin{bedi}}
\newcommand{\ebd}{\end{bedi}}

\newtheorem{bedin}[definition]{{\sc Conditions}\sc}
\newcommand{\bbdn}{\begin{bedin}}
\newcommand{\ebdn}{\end{bedin}}

\newtheorem{corollary}[definition]{{\sc Corollary}\sc}
\newcommand{\bco}{\begin{corollary}}
\newcommand{\eco}{\end{corollary}}

\newtheorem{lemma}[definition]{{\sc Lemma}\sc}
\newcommand{\blem}{\begin{lemma}}
\newcommand{\elem}{\end{lemma}}

\newtheorem{proposition}[definition]{{\sc Proposition}\sc}
\newcommand{\bpro}{\begin{proposition}}
\newcommand{\epro}{\end{proposition}}

\newtheorem{satz}[definition]{{\sc Theorem}\sc}
\newcommand{\bsa}{\begin{satz}}
\newcommand{\esa}{\end{satz}}

\newtheorem{theorem}[definition]{{\sc Theorem}\sc}
\newcommand{\bth}{\begin{theorem}}
\newcommand{\eth}{\end{theorem}}

\newtheorem{assumption}[definition]{{\sc Assumption}\sc}
\newcommand{\bas}{\begin{assumption}}
\newcommand{\eas}{\end{assumption}}

\newtheorem{assumptions}[definition]{{\sc Assumptions}\sc}
\newcommand{\bass}{\begin{assumptions}}
\newcommand{\eass}{\end{assumptions}}

\newtheorem{abb}{{\sc Figure}\sc}
\newcommand{\babb}{\begin{abb}}
\newcommand{\eabb}{\end{abb}}

\newenvironment{remark}{\begin{rmk}\sl}{\end{rmk}}
\newtheorem{rmk}{{\sc Remark}\sc}[section]
\newcommand{\brem}{\begin{remark}}
\newcommand{\erem}{\end{remark}}

\newenvironment{remarks}{\begin{rmks}\sl}{\end{rmks}}
\newtheorem{rmks}{{\sc Remarks}\sc}[section]
\newcommand{\brems}{\begin{remarks}}
\newcommand{\erems}{\end{remarks}}

\newenvironment{example}{\begin{exmp}\rm}{\end{exmp}}
\newtheorem{exmp}{{\sc Example}\sc}[section]
\newcommand{\bbsp}{\begin{example}}
\newcommand{\ebsp}{\end{example}}
\newcommand{\bexa}{\begin{example}}
\newcommand{\eexa}{\end{example}}

\newtheorem{model}{{\sc Model}\sc}[section]
\newcommand{\bmdl}{\begin{model}}
\newcommand{\emdl}{\end{model}}

\newtheorem{scheme}{{\sc Scheme}\sc}[section]
\newcommand{\bscm}{\begin{scheme}}
\newcommand{\escm}{\end{scheme}}

\newenvironment{tabelle}{\begin{tabl}\sl}{\end{tabl}}
\newtheorem{tabl}{{\sc Table}\sc}
\newcommand{\btab}{\begin{tabelle}}
\newcommand{\etab}{\end{tabelle}}

\newenvironment{exercise}{\begin{exc}\sl}{\end{exc}}
\newtheorem{exc}{Exercise}[section]
\newcommand{\bexe}{\begin{exercise}}
\newcommand{\eexe}{\end{exercise}}

\newcommand{\db}{distribution}

\newcommand{\dbs}{dis\-tri\-bu\-tions}

\newcommand{\yp}{hypothesis}

\newcommand{\np}{non\-pa\-ra\-me\-tric}
\newcommand{\stat}{statistic}

\newcommand{\ci}{confidence interval}
\newcommand{\cis}{confidence intervals}

\newcommand{\asy}{asymptotic}

\newcommand{\trt}{treatment}
\newcommand{\trts}{treatments}
\newcommand{\Trt}{Treatment}

\newcommand{\ig}{i=1, \ldots,}

\newcommand{\logit}{\operatorname{\it logit}}

\newcommand{\nfrac}{\nicefrac}

\usepackage{nicefrac}

\newcommand{\hlam}{{\widehat \lambda}}

\newcommand{\htheta}{{\widehat \theta}}

\relax

\newcommand{\lamso}{\lam_{\text{SO}}}
\newcommand{\lamwr}{\lam_{\text{WR}}}
\newcommand{\hlamso}{\hlam_{\text{SO}}}
\newcommand{\hlamwr}{\hlam_{\text{WR}}}

\newcommand{\lamwmw}{\lam_{\text{WMW}}}
\newcommand{\ollamso}{\ollam_{\text{SO}}}
\newcommand{\ollamwr}{\ollam_{\text{WR}}}

\newcommand{\olx}{\overline{x}}
\newcommand{\oly}{\overline{y}}

\newcommand{\olD}{\overline{D}}

\newcommand{\ollam}{\overline{\lambda}}

\newcommand{\oltheta}{\overline{\theta}}

\begin{document}

% wrso-int-HH.tex

\begin{titlepage}
\title{\bf \Huge \bf The Success-Odds - A Modified Win-Ratio} 
\author{Edgar Brunner \\
University of Göttingen \\
Institut für Medizinische Statistik \\[50ex]
{\it English version of the German handout to the talk} \\ ''Win-Ratio and 
Mann-Whitney Odds'' \\ {\it at the Fall Workshop on Statistical Methods in 
Medical Research} \\ {\it of the IBS / DR in Hamburg} \\ 
{\it 21.-22. November, 2019} \\ 
{\it https://www.unimedizin-mainz.de/smde/herbstworkshop-2019.html }} 

\thispagestyle{empty}
\maketitle 
\end{titlepage}

% wrso-int-HH.tex

%%%%% Überarbeitet: 19.2.2020 EB %%%%%%%%%%%%%%%%%%%%%%%%%%%%%%%%%%%%%

\section{Introduction} 

Multiple and combined endpoints involving also non-normal outcomes appear in 
many clinical trials in various areas in medicine where the outcome may be 
observed not only on a metric scale. In some cases, the outcome can be observed 
only on an ordinal or even dichotomous scale. Then the success of two 
therapies then can only be assessed by comparing the outcome of two arbitrary 
selected patients from the two therapy groups by {\it 'better'}, {\it 'equal'} 
or {\it 'worse'}. Now let 
$X \sim F_1(x)$ denote the outcome of therapy $A$ and $Y \sim F_2(x)$ denote the 
outcome of therapy $B$. Then, for the three potential results 
\ben
 \item[(1)] \ $X>Y$ ($A$ better than $B$), 
 \item[(2)] \ $X=Y$ ($A$ equal or comparable to $B$), 
 \item[(3)] \ $X<Y$ ($A$ worse than $B$) 
\een
these outcomes can be quantified by the three probabilities $p^- = P(X<Y), \ 
p_0 =  P(X=Y)$, and $p^+ = P(X>Y)$, where $p^- + p_0 + p^+ = 1$. The outcomes 
$X$ and $Y$ can be measured or observed on an appropriate metric or ordinal 
scale.

To compare the underlying \dbs\ $F_1$ and $F_2$, the Mann-Whitney test (1947) 
is established since many decades. To test the \yp\ $H_0: F_1=F_2$ using the 
effect $p^+ = P(X>Y)$, this test had been developed for the case of continuous 
\dbs, i.e. for the case of no ties where $p_0=0$. The original Mann-Whitney 
test is consistent to alternatives of the form $p^+ \neq \nfrac12$. Later, 
Putter (1955) considered the case where also ties are admitted ($p_0>0$) and 
showed that this modified test is based on the quantity 
\bqan
 \theta &=& p^+ + \tfrac12 p_0 \ = \ P(X>Y) + \tfrac12 P(X=Y) \label{releff} 
\eqan
and is consistent to alternatives of the form $\theta \neq \nfrac12$. Giving 
credit to Wilcoxon (1945, 1947), this test is also called Wilcoxon-Mann-Whitney 
test (WMW-test). The quantity $\theta$ can be well interpreted as the 
probability that therapy $A$ is better than $B$ (plus $\nfrac12$-times the 
probability that the two therapies are comparable). For the clinician, however, 
it is less comprehensible since it is not obvious which would be the benefit 
for a patient if, e.g., $\theta = 0.667$. For this reason, Noether (1987) 
introduced the effect $\lam = P(X>Y) / P(X<Y) = p^+ / p^-$ as a well 
comprehensible effect assuming continuous \dbs\ ($p_0=0$). Unfortunately, the 
quantity $\lam$ in this paper, had been denoted as 'odds-ratio' although these 
are odds $\lam = \theta / (1-\theta)$ since $p^+ = 1-p^-$ for $p_0=0$. 
Moreover, as this paper appeared in a  more theoretically oriented journal, 
this quantity has not been perceived by the practitioners and clinicians. 

Fortunately, this idea was seized again by Pocock et al. (2012) as an intuitive 
and well comprehensible effect and was denoted as {\it 'win-ratio'} (WR)
\bqan 
\lamwr &=& P(X>Y) / P(X<Y) \ = \ p^+ / p^-.  \label{lamwrdef}
\eqan

Later, this quantity had been suggested by Wang und Pocock (2016) also for 
general non-normal outcomes in clinical trials. Unlike Noether (1987), Wang and 
Pocock (2016), however, explicitly allowed for ties in the data. This means 
that $p_0>0$ without including the term $p_0$ in the definition of the quantity 
$\lamwr$. Motivated by the consideration of effects for ordinal data, O'Brien 
and Castelloe (2006) suggested the quantity
\bqa 
\lamwmw &=& \frac{P(X>Y) + \frac12 P(X=Y)}{P(X<Y) + \frac12 P(X=Y)} \ = \ 
               \frac{\theta}{1-\theta} 
\eqa
as a well interpretable effect but did not consider this quantity in more 
detail. Later, Dong et al. (2019) discussed a \stat\ $\text{WO} = U_2/(1-U_2)$, 
where $U_2 = \htheta_N$ denotes the estimator of the Mann-Whitney effect $p^+ = 
P(X>Y)$ in its generalized version $\theta = p^+ + \frac12 p_0$ including the 
case of ties (Putter, 1955). Since that time, the quantity $\theta$ got many 
divers denominations in the different areas of applications. For continuous 
\dbs\ $F_1$ and $F_2$, Birnbaum and Klose (1957) considered the function $L(t) 
= F_1\left[ F_2^{-1}(t) \right]$, which they denoted as a {\it 'relative \db\ 
of $X$ and $Y$'}. Since $p^+ = P(X>Y)$ is the expectation of $L(t)$, i.e., $p^+ 
= \int_0^1 t dL(t)$, it is called {\it 'relative effect'} with regard to 
Birnbaum and Klose (1957), see for example, Brunner and Puri (1996, 2001) and 
references cited therein. This terminology points out that $p^+$ describes an 
effect of $F_1(x)$ with respect to $F_2(x)$. Its extension $\theta = p^+ + 
\frac12 p_0$, which is also valid in case of ties reduces to $p^+$ for 
continuous \dbs\ since $p_0 = 0$ in this case.

When comparing two therapies $A$ and $B$, a success of $A$ in relation to $B$ 
can be described by the probability $\theta = p^+ + \frac12 p_0$, where $\theta 
> \nfrac12$ means a success of $A$ over $B$. Then the quantity $\theta / (1 - 
\theta)$ is the chance to obtain a better result applying $A$ instead of $B$. 
Therefore it shall be called {\it success odds} (SO) and is denoted by 
\bqan
\lamso &=& \theta /(1-\theta) \ = \ \frac{p^+ + \frac12 p_0}{p^- + \frac12 p_0} 
\label{lamsodef}
\eqan
relating a success $\theta > \nfrac12$ to the success-odds $\lamso > 1$. 
Basically, it is a simple modification of the win-ratio $p^+/p^-$ by adding 
half of the probability of ties, $p_0 = P(X=Y)$, to the numerator and the 
denominator extending the win-ratio (and in turn Noether's ratio) $p^+ / p^-$ 
to the case of ties. Note that $\theta$ quantifies the \np\ effect of the 
WMW-test in case of ties and the consistency region of this test is given by 
$\theta \neq \nfrac12$.

It is the aim of this manuscript to investigate the properties of $\lamwr$ and 
$\lamso$ in case of ties since they are included in the definition of $\lamso$ 
but not in the definition of $\lamwr$ .

%wrso-twotr-HH.tex

%%%%% Überarbeitet: 20.1.2020 EB %%%%%%%%%%%%%%%%%%%%%%%%%%%%%%%%%%%%%

\section{Comparison of two Treatments} 
\subsection{Illustration of $P(X<Y)=0$} \label{invalid}

First we consider the simple case of two treatments $A$ and $B$ as explained in 
the introduction. In general, it holds for $P(X>Y) \ge P(X<Y)$ and $p_0 = 
P(X=Y) \ge 0$ that
\bqan
 \lamwr \ = \ \frac{P(X>Y)}{P(X<Y)} = \frac{p^+}{p^-} & \ge & 
\frac{p^+ + \frac12 p_0}{p^- + \frac12 p_0} \ = \ \lamso \ , \label{lamwrso}
\eqan
where $P(X<Y) > 0$ must be assumed. Equality in (\ref{lamwrso}) holds if and 
only if
\ben
 \item[(1)] \ either \ $p_0 = 0$ \ (i.e. no ties) 
 \item[(2)] \ or \ $p^+ = p^-$ \ ($\lamwr = \lamso =1$). 
\een

Thus, in all other cases, $\lamwr > \lamso$, by definition. In the sequel, the 
impact of ties on the WR $\lamwr$ and on the SO $\lamso$ shall be demonstrated 
by means of some examples. In the first example it is demonstrated that $P(X<Y) 
=0$ invalidates the WR $\lamwr$ but not the SO $\lamso$.

%\subsection{Paarweise Vergleiche von 3 \Bhn}
\bexa (Pairwise comparisons of 3 treatments) \ \label{pm0}
In this example, three \dbs\ $F_1, F_2$ and $F_3$ defined on an ordinal scale 
are compared. The ordinal categories are labeled by 1, 2 and 3 where the result 
$x=3$ is better than the results $x=2$ or $x=1$ and the result $x=2$ is better 
than $x=1$. Let $F_1$ denote the \db\ of the result for treatment $A$, $F_2$ 
for treatment $B$, and $F_3$ for treatment $C$. The probabilities $f_i$ for the 
results $x=1, x=2$ and $x=3$ of the discrete \dbs\  $F_i$, $i=1,2,3$ are 
displayed in Table~\ref{pm0}.
\btab \label{pm0} 
The results for the treatments $A, B$, and $C$ are described by the \dbs\ $F_1, 
F_2$, and $F_3$ with probabilities $f_1$, $f_2$ and $f_3$ for the discrete 
outcomes $1, 2$, and $3$.
\bcen
 \btb{|c|l|l|l|l|} \hline
 Treatment & Probabilities & $x=1$ & $x=2$ & $x=3$ \\ \hline
     $A$    & $f_1(x)$ & $0.10$  & $0.90$  & $0$     \\
		 $B$    & $f_2(x)$ & $0$     & $0.90$  & $0.10$  \\
		 $C$    & $f_3(x)$ & $0$     & $0.10$  & $0.90$  \\ \hline
 \etb
\ecen		
\etab

The values of the relative effect $\theta$, as well as of the effects $\lamwr$ 
and $\lamso$ for the pairwise comparisons of the treatments $A, B$, and $C$ are 
listed in Table~\ref{pm0ij}.
\btab \label{pm0ij}
Pairwise comparisons of the treatments $A, B$, and $C$ by means of the related 
relative effect $\theta$, the SO $\lamso$, and the WR $\lamwr$. 
\bcen
 \btb{|c|l|c|r|} \hline
  Comparison & \mc{1}{c|}{$\theta$} & $\lamwr$ & $\lamso$ \\ \hline
   $B,A$  & $0.595$         & $\infty$       &  $1.47$ \\
	 $C,B$  & $0.900$         & $81$           &  $9.00$ \\ 
	 $C,A$  & $0.955$         & $\infty$       & $21.22$ \\ \hline
 \etb
\ecen	
\etab

Obviously, it can be seen from Table~\ref{pm0} that treatment $C$ is much 
better than treatment $A$ and also better than treatment $B$ while treatment 
$B$ is slightly better than $A$. This is well characterized by the relative 
effect $\theta$ and by the SO $\lamso$ while the WR $\lamwr$ is not able to 
reasonably describe the successes of the treatments. It shall be noted that in 
the present example the pairwise comparisons cannot lead to non-transitive 
decisions since the three \dbs\ are stochastically ordered. This is immediately 
seen from Table~\ref{pm0} where $F_3(x) \le F_2(x) \le F_1(x)$ for all $x$. 
\eexa

\subsection{Metric and Ordinal Data} 

\bexa \label{rounding} (Coarsening of the meaurement scale) \ It shall be 
demonstrate by this example that a coarsening of the measurement can lead to an 
increase of the WR while the relative effect $\theta$ and SO $\lamso$ may 
remain unchanged. When coarsening the measurement scale, the means in case of 
metric data and in turn their differences as well as the relative effects may 
change. Therefore, the \dbs\ in this example are chosen in such a way that the 
means $\olx_{i \cdot}$ in \trt\ $A$ and $\oly_{i \cdot}$ in \trt\ $B$ remain 
unchanged in the three steps of the coarsening. In the same way, the relative 
effects $\theta_i = P(X_i > Y_i) + \frac12 P(X_i = Y_i)$, $\ig 4$, remain 
unchanged which implies that the SO $\lamso(i)$ remain also unchanged. The 
proportion of the ties, however, increases in the three steps of the coarsening 
which leads to an increase of the WR $\lamwr(i)$. The coarsening of the 
measurements in the three steps was performed by rounding the measurements in 
the following table. 
\btab \label{roundproc} Description of rounding measurements in three steps. 
\\[1ex]
\btb{ll}
Case (1) & The measurements are observed with an accuracy of one place after 
           the decimal \\
         & point. \\ 
Case (2) & The measurements are rounded to integers. \\
Case (3) & The measurements within the interval $[2.6, 4.4]$ are rounded to the 
           mean 3.5 \\
				 & of this interval while the other values remained integers. \\
Case (4) & The measurements within the interval $[1.6, 5.4]$ are rounded to the 
           mean 3.5 \\ 
				 & of this interval while the other values remained integers.
\etb
\etab

Neither in a parametric model nor in a \np\ model different \trt\ effects are 
obtained since the means in the \trts\ $A$ and $B$ - and in turn the 
differences - as well as the relative effects remained the same in the three 
steps of the coarsening. Thus, the SO $\lamso(1) = \cdots = \lamso(4) = 
2.125$ are identical in all steps. The WR, however, increases from $\lamwr(1) = 
2.125$ to $\lamwr(3) = 2.8$ and becomes $\lamwr(4) = \infty$ in the last step. 
The measurements and their coarsening are listed in Table~\ref{tabround} along 
with the means for the \trts\ $A$ and $B$.
\bcen
\btab \label{tabround}
Measurements for the \trts\ $A$ and $B$ (first row) and the same measurements 
rounded as described above (rows $2 - 4$). 
\bcen
\btb{|c|lllll||lllll||c|c|} \hline
 \mc{1}{|c}{ } & \mc{10}{c||}{Measurements} & \mc{2}{c|}{Menas} \\ \hline
Case & \mc{5}{c||}{\Trt\ \ $A$ \ $(x_1, \ldots, x_5)$} & \mc{5}{c||}{\Trt\ $B$ 
\ $(y_1, \ldots, y_5)$} & \ $A$ \ & \ $B$ \ \\ 
 \hline
 1 & 1.7 & 3.3 & 3.8 & 4.9 & 6.3 & 1.4 & 1.6 & 2.7 & 4.3 & 5.0 &  4 & 3 \\ 
 2 & 2 & 3 & 4 & 5 & 6 & 1 & 2 & 3 & 4 & 5 &  4 & 3 \\ 
 3 & 2 & 3.5 & 3.5 & 5 & 6 & 1 & 2 & 3.5 & 3.5 & 5 &  4 & 3 \\ 
 4 & 3.5 & 3.5 & 3.5 & 3.5 & 6 & 1 & 3.5 & 3.5 & 3.5 & 3.5 &  4 & 3 \\ \hline
\etb
\ecen
\etab
\ecen

The proportion of ties $p_0$, the differences, relative effects $\theta$, SO 
$\lamso$ and the WR $\lamwr$ are listed in Table~\ref{resround}.
\bcen
\btab \label{resround}
Changes of the WR $\lamwr$ for the comparison of the \trts\ $A$ and $B$ when 
coarsening the measurement scale where the proportion of ties $p_0 = P(X_i = 
Y_i)$ is increased while the means as well as the relative effects remain 
unchanged. \\[1ex]
\bcen
\btb{|c|c|c|c|c|l|} \hline
Case & $p_0$ & Diff. & Relative Effect & SO $\lamso$ & WR $\lamwr$ \\ \hline
1 & 0.00 & 1 & 0.68 & 2.125 & \qquad 2.125 \\ 
2 & 0.16 & 1 & 0.68 & 2.125 & \qquad 2.5 \\ 
3 & 0.24 & 1 & 0.68 & 2.125 & \qquad 2.8 \\ 
4 & 0.64 & 1 & 0.68 & 2.125 & \qquad $\infty$ \\ \hline
\etb
\ecen
\etab
\ecen
\eexa 

\bexa \label{ordinal} (Combining ordinal categories) \ In this example it is 
demonstrated how the WR $\lamwr$ might change if in an ordinal scale involving 
6 ordinal categories 1, 2, 3, 4, 5, 6 the three categories 3, 4, 5 are 
combined in a new category 4. The relative effect $\theta$ and the SO $\lamso$ 
remain unchanged in this case.

The probabilities $f_i(A)$ of the results $X = i$ (\Trt\ $A$) and $f_i(B)$ of 
the results $Y = i$ (\Trt\ $B$), $\ig 6$, are displayed in the upper part of 
Table~\ref{ordinaldat1}, the proportion of ties $p_0=P(X=Y)$, the relative 
effect $\theta$, the SO $\lamso$ as well as the WR $\lamwr$ are displayed in 
the lower part of Table~\ref{ordinaldat1}. It may be noted that here, $\lamwr > 
\lamso$ by definition since $p_0 = 0.13 > 0$ according to the explanations in
Section~\ref{invalid}. 
\bcen
\btab \label{ordinaldat1}
Probabilities for the ordinal scores 1 to 6 for the two \trts\ $A$ and $B$, the 
proportion of ties $p_0$, the relative effect $\theta$, the SO $\lamso$, and 
the WR $\lamwr$. \\[1ex]
\bcen
\btb{|c|c|c|c|c|c|c|} \hline
& \mc{6}{|c|}{Score} \\ \cline{2-7}
\Trt\ & 1 & 2 & 3 & 4 & 5 & 6 \\ \hline
$A$ &  0.0 & 0.1 & 0.2 & 0.3 & 0.2 & 0.2 \\
$B$ &  0.3 & 0.3 & 0.1 & 0.2 & 0.1 & 0.0 \\ \hline \hline
$p_0=P(X=Y)$ & \mc{2}{c|}{$\theta$} & \mc{2}{c|}{$\lamso$} & 
\mc{2}{c|}{$\lamwr$} \\ \hline
0.13 & \mc{2}{c|}{0.805} & \mc{2}{c|}{4.13} & \mc{2}{c|}{5.69} \\ \hline
\etb
\ecen
\etab
\ecen

The probabilities $f_i(A)$ for the results $X = i$ (\trt\ $A$) and $f_i(B)$ for 
the results $Y = i$ (\trt\ $B$), $i=1,2,4,6$, are listed in the upper part of 
Table~\ref{ordinaldat2}. Here, the categories 3, 4, 5 are combined to a new 
category 4. In the lower part of Table~\ref{ordinaldat2}, the proportion of 
ties $p_0=P(X=Y)$, the relative effect $\theta$, SO $\lamso$ and WR $\lamwr$ 
are listed for the new categories. Compared with Table~\ref{ordinaldat1}, the 
proportion of ties increased from $13\% $ to $31\% $ while the relative effect 
$\theta$ and in turn the SO $\lamso$ remained unchanged but the WR $\lamwr$ 
increased from $5.69$ to $16.25$.
\bcen
\btab \label{ordinaldat2}
Probabilities of the combined ordinal scores 1, 2, 4, 6 for the two \trts\ $A$ 
and $B$ as well as the proportion of ties $p_0$, the relative effect $\theta$, 
the SO $\lamso$, the WR $\lamwr$. \\[1ex]
\bcen
\btb{|c|c|c|c|c|} \hline
& \mc{4}{|c|}{Score} \\ \cline{2-5}
\Trt\ & 1 & 2 & 4 & 6 \\ \hline
$A$ &  0.0 & 0.1 & 0.7 & 0.2 \\
$B$ &  0.3 & 0.3 & 0.4 & 0.0 \\ \hline \hline
$p_0=P(X=Y)$ & $\theta$ & \mc{2}{c|}{$\lamso$} & $\lamwr$ \\ \hline
0.31 & 0.805 & \mc{2}{c|}{4.13} & 16.25 \\ \hline
\etb
\ecen
\etab
\ecen
\eexa

\subsection{Dichotomuous Data} \label{dicho}

In case of binary data for the treatments $A$ and $B$ with success 
probabilities $q_A = P(X=1)$ and $q_B = P(Y=1)$ the quantities WR and SO are 
given by 
\bqa
 \lamwr \ = \ \frac{q_A (1-q_B)}{q_B (1-q_A)} \sep \text{and} \sep \lamso \ = \ 
     \frac{q_A(1-q_B) + p_0/2}{q_B(1-q_A) + p_0/2} \ ,
\eqa
where $p_0 = \left[q_A q_B + (1-q_A)(1-q_B) \right]$ and thus by definition, 
$\lamso < \lamwr$. In this particular case, $\lamso$ may be considerably 
smaller than $\lamwr$ which, in case of dichotomous data, equals the well-known 
odds-ratio 
\bqa
 \text{OR}(A,B) &=& \frac{q_A}{1-q_A} \Big/ \frac{q_B}{1-q_B}	\ = \ 
       \frac{q_A(1-q_B)}{q_B(1-q_A)},
\eqa
which is the ratio of the success rates of both treatments $A$ and $B$ while 
$\lamso$ is based on the well-accepted Mann-Whitney effect $\theta$ (relative 
effect) in its generalized form (Putter, 1955) which includes the case of ties.  

\bexa \label{exabernoulli}
The aim of this example is to investigate whether the Win-Ratio $\lamwr$ (or 
the Odds-Ratio OR) and the Success-Odds $\lamso$ are intuitive and  well 
interpretable quantities to describe a treatment effect of a therapy $A$ with 
respect to a therapy $B$ in case of dichotomous data. The success rates $q_A = 
P(X=1)$ and $q_B = P(Y=1)$ as well as the success failures $1-q_A$ and $1-q_B$ 
are displayed in Figure~\ref{bernoulli}. %\\[-2ex]
\bcen 
%\hspace*{7ex} 
\bmp[b]{0.3\textwidth} 
 \unitlength2.1ex
\bpic(18,12)
\thicklines
 \put(0,0){\vector(1,0){12}}
 \put(0,0){\vector(0,1){10}}
 \put(0,8){\line(-1,0){0.5}}
 \put(-1.4,7.6){\parbox[b]{2ex}{\large $1$}}

 \linethickness{0.05ex}
 \put(0,8){\line(1,0){10}}
\thicklines
 \put(2,0){\line(0,-1){0.5}}
 \put(1.65,-1.6){\parbox[b]{2ex}{\large $0$}}
 \put(9,0){\line(0,-1){0.5}}
 \put(8.65,-1.6){\parbox[b]{2ex}{\large $1$}}
 \linethickness{1ex}
{\color{blue}
 \put(1.5,0){\line(0,1){1.44}}
}
{\color{red}
 \put(1.8,0){\line(0,1){3.2}}
}
 \put(8.5,6){\parbox[b]{8ex}{\tiny $q_A=0.82$}}
 \put(9,4.3){\parbox[b]{8ex}{\tiny $q_B=0.6$}}
{\color{blue}
 \put(8.0,0){\line(0,1){6.4}}
}
{\color{red}
 \put(8.2,0){\line(0,1){4.8}}
}
\put(-0.6,1.8){\parbox[b]{8ex}{\tiny $1-q_A$ \\ $=0.18$}}
 \put(2.1,2.9){\parbox[b]{8ex}{\tiny $1-q_B=0.4$}}
\put(0.5,11){\parbox[b]{12ex}{$\lamwr=3.06$}}
\put(0.5,9.5){\parbox[b]{12ex}{$\lamso=1.57$}}
\epic
\emp
\hspace*{8ex}
\bmp[b]{0.3\textwidth} 
 \unitlength2.2ex
 \bpic(18,12)
\thicklines
 \put(0,0){\vector(1,0){12}}
 \put(0,0){\vector(0,1){10}}
 \put(0,8){\line(-1,0){0.5}}
 \put(-1.4,7.6){\parbox[b]{2ex}{\large $1$}}
\linethickness{0.05ex}
 \put(0,8){\line(1,0){10}}
\thicklines
 \put(2,0){\line(0,-1){0.5}}
 \put(1.65,-1.6){\parbox[b]{2ex}{\large $0$}}
 \put(9,0){\line(0,-1){0.5}}
 \put(8.65,-1.6){\parbox[b]{2ex}{\large $1$}}
\linethickness{1ex}
{\color{blue}
 \put(1.65,0){\line(0,1){0.15}}
}
{\color{red}
 \put(1.68,0){\line(0,1){0.5}}
}
{\color{blue}
 \put(8.0,0){\line(0,1){7.8}}
}
{\color{red}
 \put(8.2,0){\line(0,1){7.6}}
}
\put(-1,0.3){\parbox[b]{8ex}{\tiny $1-q_A$ \\ $=0.01$}}
 \put(1.7,0.3){\parbox[b]{8ex}{\tiny $1-q_B=0.03$}}
\put(4.6,7.5){\parbox[b]{8ex}{\tiny $q_A=0.99$}}
 \put(8.6,7.2){\parbox[b]{8ex}{\tiny $q_B=0.97$}}

\put(1,10.8){\parbox[b]{12ex}{$\lamwr=3.06$}}
\put(1,9.5){\parbox[b]{12ex}{$\lamso=1.04$}}
\epic
\emp
\ecen
\text{ } \\[-3ex]
\babb \label{bernoulli}
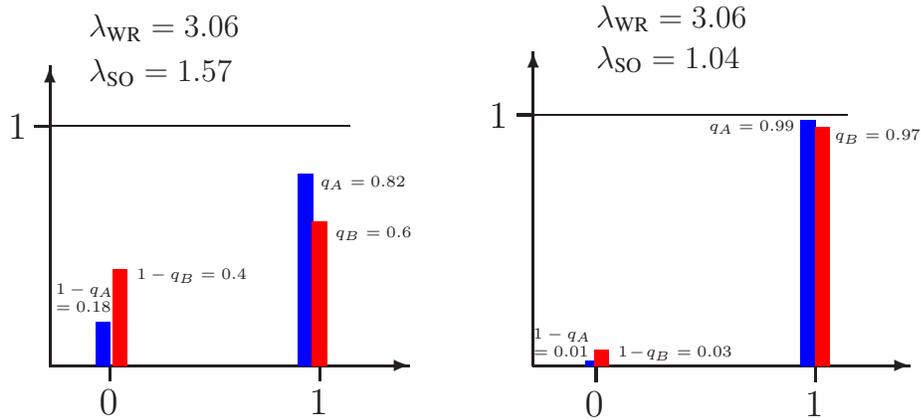
The results of the two treatments $A$ and $B$ with dichotomous endpoints are 
displayed in the two graphs. The success probabilities are $q_A=0.821$ and 
$q_B=0.6$ in the left-hand graph and $q_A=0.99$ and $q_B=0.97$ int the 
right-hand graph. Obviously, in the left-hand graph a clear difference of the 
successes of both therapies can be seen while in the right-hand graph nearly no 
difference can be recognized between the two treatments. Moreover, in the 
right-hand graph about $96\% $ of the results (or more precisely, $q_A \cdot 
q_B + (1-q_A)(1-q_B)=0.961$) are identical. These circumstances, however, are 
not depicted by the Win-Ratio $\lamwr$ since in both cases, $\lamwr = 3.06$. In 
contrast, the Success-Odds $\lamso$ intuitively depicts this actual situation 
since $\lamso = 1.57$ in the left graph is larger than $\lamso = 1.04$ in the 
right graph. 
\eabb
\eexa

It appears that the win-ratio $\lamwr$ does neither provide an intuitive and 
well interpretable quantification of a treatment effect for dichotomous data
nor it depicts an intuitive therapy success of therapy~$A$ over therapy~$B$.
In the sequel this is demonstrated by another example involving dichotomous 
data.

\bexa \label{rates9095}
Consider the case where the success of therapy~$A$ is increased from $q_A = 
90\% $ to $q_A=95\% $ while the therapy success of therapy $B$ is kept fixed. 
Moreover, the percentage of ties $p_0 = P(X=Y)$ remains nearly constant when 
$q_B$ ist fixed. The results are listed in the following table.
\btab \label{vergleich}
Comparison between the Win-Ratio $\lamwr$ and the Success-Odds $\lamso$ to 
intuitively depict a superiority of therapy $A$ over therapy $B$. 
\bcen
\btb{|l|c|l|r|c|} \hline
$q_A$ & $q_B$ & $p_0=P(X=Y)$ & $\lamwr=\text{OR}(A,B)$ & $\lamso = 
\frac{\theta}{1-\theta}$ \\ \hline
0.9  & 0.5 & \hspace*{5ex} 0.5  &  9.0 \hspace*{5ex} \text{ }& 2.3 \\
0.95 & 0.5 & \hspace*{5ex} 0.5  & 19.0 \hspace*{5ex} \text{ }& 2.6 \\ 
\hline \hline
0.9  & 0.6 & \hspace*{5ex} 0.58 &  6.0 \hspace*{5ex} \text{ }& 1.9 \\
0.95 & 0.6 & \hspace*{5ex} 0.59 & 12.7 \hspace*{5ex} \text{ }& 2.1 \\ 
\hline \hline
0.9  & 0.7 & \hspace*{5ex} 0.66 &  3.9 \hspace*{5ex} \text{ }& 1.5 \\
0.95 & 0.7 & \hspace*{5ex} 0.68 &  8.1 \hspace*{5ex} \text{ }& 1.7 \\ \hline
\etb
\ecen
\etab

It appears from Table~\ref{vergleich} that the Win-Ratio $\lamwr$ is 
approximately doubled independently of the success rate of therapy $B$ if the 
success rate $q_A$ of therapy $A$ is slightly increased from $90\% $ to 
$95\% $. In a graphical representation, this difference would hardly be 
recognized. 

In conclusion, it appears that in the case of dichotomous data, the win-ratio 
$\lamwr$ looses its appealing property to provide an intuitive quantification 
of a therapy effect as a chance to obtain a better result by applying 
therapy~$A$ instead of therapy~$B$.
\eexa

In the next section, the conclusions from the examples presented in the 
previous sections shall be summarized anf discussed.

\subsection{Discussion of the Win-Ratio for Two Samples} \label{disc}

Basically, the idea of the win-ratio $\lamwr$ to provide an intuitive and 
well-interpretable effect when the result of a therapy can only be assessed by 
'better', 'worse' or 'comparable', is to be welcomed. However, the proportion 
of ties (comparable results) must be included in it's definition since ties are 
allowed in the model. Otherwise, this quantity has some annoying properties.
\ben
 \item The computation of the WR $\lamwr$ breaks down if $P(X<Y)=0$ while the 
       SO $\lamso$ depicts this case also and can only break down in the case 
			 where $P(X<Y)=0$ and $P(X=Y)=0$, i.e. in the trivial case of a one-point 
			 \db\ (see the discussion in Section~\ref{invalid}).
 \item It is counterintuitive that an effect can increase if the measurements 
       are less precise or the data are observed less accurately. This is 
			 demonstrated in Examples~\ref{rounding} and \ref{ordinal}. Also such a 
			 property would offer a possibility to manipulations.
 \item In case of dichotomous data, the win-ratio $\lamwr$ looses its appealing 
       property to provide an intuitive quantification of a therapy effect in 
			 general. This, however, was the basic idea of the win-ratio. An example 
			 is discussed in Section~\ref{dicho}.
\een

Thus, the nice idea of the win-ratio should only be used in it's modified or 
improved form of the success-odds $\lamso$ which appeared in the literature 
already in the conference paper by O'Brien and Castelloe (2006) - unfortunately 
without any further discussion. It extends Noether's idea to provide an 
intuitive treatment effect for the Mann-Whitney test to the case of ties. Also, 
Dong et al. (2019) as well as Gasparyan and Koch (2019) consider the 
success-odds $\lamso$ but did not discuss the drawbacks of the win-ratio 
$\lamwr$ in case of ties. They have first been considered in detail in the talk 
by Brunner (2019) at the fall-workshop of the working-group 'Statistical 
Methods in Medical Research' of the IBS / DR in Hamburg on November, 22, 1019. 

In summarizing this discussion, the success-odds $\lamso$ can be recommended as 
an improved version of the win-ratio. Therefore, the next section briefly 
discusses tests and \cis\ for the success-odds $\lamso$ and - for completeness -
also for the win-ratio $\lamwr$.

\iffalse
the \asy\ distribution of the estimator $\hlamso 
= \htheta / (1 - \htheta)$, where $\htheta$ denotes the numerator of the 
Wilcoxon-Mann-Whitney \stat. To derive a test for the \yp\ $H_0: \lamso = 1$, 
the \asy\ \db\ of $\htheta$ is not only considered under the \yp\ but also in 
the general case under alternatives. Also \cis\ for $\lamso$ are considered 
using the well-known results of the Wilcoxon-Mann-Whitney \stat\ in the 
so-called \np\ Behrens-Fisher setting (e.g., Brunner and Munzel, 2000 or the 
recent book by Brunner et al., 2019, Section~3).
\fi

\subsection{Tests and Confidence Intervals for $\lamwr$ and $\lamso$}	
\label{asyv}
\subsubsection{Win-Ratio $\lamwr$}
The \asy\ \db\ of $\hlamwr$ and \cis\ for $\lamwr$ have been derived by Bebu 
and Lachin (2016) and by Dong, Ballerstedt, and Vandemeulebroecke (2016) where 
also R- and SAS-programs to perform the computations are provided.

\subsubsection{Success-Odds $\lamso$}
Estimators for the relative effect $\theta$ in (\ref{releff}) are available 
from the literature. A test of the \yp\ $H_0^\theta: \theta = \nfrac12$ in a 
general model including also the case of ties is considered by Brunner and 
Munzel (2000), for example. This is known as the \np\ Behrens-Fisher Problem. 

It may be noted that the \yp\ $H_0^\theta: \theta = \nfrac12$  is equivalent to 
$H_0^\lam: \lamso = 1$. For more details we refer to Section~3.5 of the 
textbook by Brunner, Bathke, and Konietschke (2019) where also a 
range-preserving \ci\ for $\theta$ is derived in Section~3.7.2. This can easily 
be extended to the success-odds $\lamso\ = \theta/(1-\theta)$ by the 
transformation $\logit(\theta)$ using Cramér's $\delta$-theorem and then 
back-transforming it to $\lamso$ by $exp(\ \cdot \ )$. The R-package 
{\it rankFD} (CRAN), which performs the computations of these quantities, is 
described in Section~A.2.2 of this book.

%wrso-threetr-HH.tex

%%%%% Überarbeitet: 20.1.2020 EB %%%%%%%%%%%%%%%%%%%%%%%%%%%%%%%%%%%%%

\section{Comparison of Several Distributions}

Pairwise comparisons using procedures based on the relative effect $\theta$ may 
lead to non-transitive decisions. This is well-known for the 
Wilcoxon-Mann-Whitney test, for example, and holds also true for the quantities 
$\lamso$ and $\lamwr$. This shall be demonstrated by the so-called tricky-dice 
(see, e.g., Peterson, 2002 or Gardner, 1970). For example, the following three 
dice \\[1ex]	
	\bcen
 	 \btb{lcccccc} 
	  D1: & 1 & 4 & 5 & 6 & 7 & 7 \\
		D2: & 3 & 3 & 4 & 5 & 6 & 9 \\
		D3: & 1 & 2 & 2 & 8 & 8 & 9 
	 \etb
  \ecen
lead to paradoxical results when pairwise comparisons are performed:	
\ben
 \item[(a)] $D1 / D2$:\sep $\theta = 0.57$, \ $\lamso = 1.32$, \ $\lamwr = 
            1.36$ \ $\RA$ \ $D1 > D2$
 \item[(b)] $D2 / D3$:\sep $\theta = 0.57$, \ $\lamso = 1.32$, \ $\lamwr = 
            1.33$ \ $\RA$ \ $D2 > D3$			
 \item[(c)] $D3 / D1$:\sep $\theta = 0.57$, \ $\lamso = 1.32$, \ $\lamwr = 
            1.33$ \ $\RA$ \ $D3 > D1$,
\een
which means that die $D1$ is better than $D2$, die $D2$ is better than $D3$, 
and that finally die $D3$ is better than $D1$. A solution of this 
non-transitivity problem might be comparing each die with a common casino-type 
die, for example a roller $D$ representing a mixture of all three dice. This is
basically the principle underlying the Kruskal-Wallis test which compares each 
\db\ with a weighted mean \db\ $D = (D1 \cup  D2 \cup D3)$. In the example 
presented above one obtains $D1 / D = D2 / D = D3 / D$ since in all cases, 
$\lamso = \lamwr = 1$. For a different common casino-type die, of course, one 
could obtain a different result.

\section{Stratified Designs}

When using a stratified version of the Wilcoxon-Mann-Whitney test, for example 
van Elteren's test (1960), a similar paradoxical decision might happen. An  
example is given in Thangavelu und Brunner (2007). This is briefly described 
below.
\bit
	  \item[ ] \btb{ccccclll}
		      & \mc{2}{c}{Therapy} & \\ \cline{2-3}
  Stratum $(i)$ & $A$  & $B$  & \sep & $\theta^{(i)}$ & $\lamso^{(i)}$ & 
	$\lamwr^{(i)}$ \\[0.4ex] \hline
	& & & & & & \\[-1.5ex]
				1 & $D1$ & $D2$ & & 0.57 & 1.32 & 1.36 \\
				2 & $D2$ & $D3$ & & 0.57 & 1.32 & 1.33 \\
				3 & $D3$ & $D1$ & & 0.57 & 1.32 & 1.33 \\ \hline
				& & & & & \\[-1.5ex]
	Means & \mc{2}{c}{$\olD_A	= \olD_B$} & & 0.57 & 1.32 & 1.34 & $\RA$ \ 
	      Therapy $A>B$ 
				\etb
\eit	

Since the means $\olD_A$ and $\olD_B$ are averaged over the same three \dbs\ on 
which the faces of the dice are based, it follows that $\olD_A	= \olD_B$ and 
thus, $\oltheta = 0.5$ and $\ollamso = \ollamwr =~1$. Thus, both the therapies 
have equal successes. The means over the stratified versions of relative 
effects $\theta^{(i)}$, the success-odds $\lamso^{(i)}$, and the win-ratios 
$\lamwr^{(i)}$ averaged over the three strata, however, demonstrate a 
superiority of therapy $A$ ($\oltheta = 0.57 > \nfrac12$, $\ollamso = 1.32 > 1$ 
and $\ollamwr = 1.34 > 1$) over therapy $B$. In some sense, this is similar to 
Simpson's paradox and is explained by the non-transitivity of the pairwise 
comparisons of the dice. Thus, different procedures must be developed for 
stratified designs which are beyond the scope of this manuscript and shall be 
discussed elsewhere.

\section{Discussion and Outlook} \label{outlook}

The idea of the win-ratio $\lamwr$ to provide a well interpretable and clear 
effect for the clinician is excellent and to be welcomed. The quantity $\lamwr$ 
as it stands, however, has some strange and undesirable properties. Thus, the 
win-ratio $\lamwr$ should be slightly modified. Such a modification $\lamso$, 
called 'success-odds' is suggested here and it has been demonstrated that 
$\lamso$ does not have the drawbacks of the the win-ratio $\lamwr$ in case of 
ties. Moreover, theoretical results are available from the literature by which 
the \asy\ \db\ of an estimator of the success-odds $\lamso$ is easily obtained.
Thus, a test of the \yp\ $H_0^\lam: \lamso = 1$ as well as a \ci\ for $\lamso$ 
can be derived using Cramér's $\delta$-theorem (see, e.g., Brunner et al., 
2019, Sections~3.5, 3.7.2, 7.4, 7.5, and 7.6.1). 

The generalization to several samples and stratified designs, however, is not 
straightforward since decisions based on $\lamso$ or $\lamwr$ may be 
non-transitive as briefly demonstrated by counter-examples in Sections~3 and 4. 
Reasonable extensions of the success-odds $\lamso$ to several samples, 
stratified and factorial designs are currently under investigation.

\section{Acknowledgment and Remarks}

The topic of this manuscript had been presented in a talk by the author at the 
fall-workshop of the working-group on 'Statistical Methods in Medicine' of the 
IBS/DR in Hamburg on November 22 in 2019. The author would like to thank the 
audience of this workshop for helpful comments and remarks. A handout in German 
language to that talk was available for the workshop. The present English 
version is based on this handout.

\newpage

% Lit-HH.tex
%%%% Überarbeitet: 28.12.19 EB %%%%%%%%%%%%%%%%%%%%%%%%%%%%%%%%%%%%%

\section{References}
\bdes 
\iffalse
\item {\sc Abdalla, S., Montez-Rath, M.E., Parfrey, P.S.} (2016). The win ratio 
approach to analyzing composite outcomes: An application to the EVOLVE trial. 
{\it Contemporary clinical trials} {\bf 48}, 119--124.

\item {\sc Akritas, M.~G., Arnold, S.~F. and  Brunner, E.} (1997). 
Nonparametric hypotheses and rank statistics for unbalanced factorial designs.
{\em Journal of the American Statistical Association} {\bf 92}, 258--265.
\fi

\item {\sc Bebu, I., Lachin, J.M.} (2016). Large sample inference for a win 
ratio analysis of a composite outcome based on prioritized components. {\it 
Biostatistics} {\bf 17}, 178--187.

\item {\sc Birnbaum, Z.~W. and Klose, O.~M.} (1957). Bounds for the Variance of 
the Mann-Whitney Statistic. {\em Annals of Mathematical Statistics} {\bf 28}, 
933--945.

\item {\sc Brunner, E.} (2019). Win-Ratio und Mann-Whitney-Odds. Talk presented 
at the fall-workshop of the working-group 'Statistical Methods in Medical 
Research' of the IBS / DR in Hamburg on November, 22, 1019. \\ 
{\it https://www.unimedizin-mainz.de/smde/herbstworkshop-2019.html }

\item {\sc Brunner, E., Bathke, A.~C., and Konietschke, F. (2019)}. {\em Rank- 
and Pseudo\--Rank Procedures in Factorial Designs - Using R and SAS - 
Independent Ob\-ser\-va\-tions.} Springer Series in Statistics, Springer, 
Heidelberg.

\item {\sc Brunner, E. and Munzel, U.} (2000). The Nonparametric Behrens-Fisher 
Problem: Asymptotic Theory and a Small-Sample Approximation. {\em Biometrical 
Journal} {\bf 42}, 17--25.

\item {\sc Brunner, E. and Puri, M. L.} (1996). Nonparametric methods in design 
and analysis of experiments. {\it Handbook of Statistics (S. Ghosh and C.R. Rao, 
Eds.)} {\bf 13}, 631--703. 

\item {\sc Brunner, E. and Puri, M. L.} (2001). Nonparametric Methods in 
Factorial Designs. {\it Statistical Papers} {\bf 42}, 1--52.

\iffalse
\item {\sc Buyse, M.} (2010) Generalized pairwise comparisons of prioritized 
outcomes in the two-sample problem. {\it Statistics in Medicine} {\bf 29}, 
3245--3257.
\fi

\item {\sc Dong, G., Li, D., Ballerstedt, S., and Vandemeulebroecke, M.} 
(2016). A generalized analytic solution to the win ratio to analyze a composite 
endpoint considering the clinical importance order among components. 
{\it Pharmaceutical Statistics} {\bf 15}, 430--437.

\item {\sc Dong, G., Hoaglin, D.C., Qiu, J., Matsouaka, R.A., Chang, Y.-W., 
Wang, J., Vandemeulebroecke, M.} (2019). The Win Ratio: On Interpretation and
Handling of Ties. {\it Statistics in Biopharmaceutical Research} {\bf }, \\
DOI: 10.1080/19466315.2019.1575279

\iffalse
\item {\sc Dong, G., Qiu, J,. Wang, D., Vandemeulebroecke, M.} (2018). 
The stratified win ratio. {\it Journal of Biopharmaceutical Statistics} 
{\bf 28}, 778-796.

\item {\sc Dong, G., Hoaglin, D.C., Qiu, J., Matsouaka, R.A., Chang, Y.-W., 
Wang, J., Vandemeulebroecke, M.} (2019). The Win Ratio: On Interpretation and
Handling of Ties. {\it Statistics in Biopharmaceutical Research} {\bf }, \\
DOI: 10.1080/19466315.2019.1575279

\item {\sc Finkelstein, D.M., Schoenfeld, D.A.} (2019). Graphing the Win Ratio 
and its components over time. {\it Statistics in Medicine} {\bf 38}, 53--61.
\fi

\item
{\sc Gardner, M.} (1970). The paradox of the nontransitive dice and the elusive 
principle of indifference. {\em Scientific  American} {\bf 223}, 110--114.

\item 
{\sc Gasparyan, S.B., Folkvaljon, F., Bengtsson, O., Buenconsejo, J. Koch. G.G.}
(2019). Adjusted Win Ratio with Stratification: Calculation Methods and 
Interpretation. arXiv:1912.09204v1 [stat.ME] 19 Dec 2019.

\iffalse
\item {\sc Govindarajulu, Z.} (1968). Distribution-free confidence bounds for 
Pr\{ X < Y \}. {\em Annals of the Institute of Statistical Mathematics} 
{\bf 20}, 229--238.

\item {\sc Lu, Y., Wang, M., Zhang, G.} (2017). A new revised version of 
McNemar's test for paired binary data. {\it Communications in Statistics - 
Theory and Methods} {\bf 46}, 10010--10024.

\item {\sc Luo, X., Tian, H., Mohanty, S., Tsai; W.Y.} (2015). An alternative 
approach to confidence interval estimation for the win ratio statistic. {\it 
Biometrics}, {\bf 71}, 139--145.

\item {\sc Luo. X., Qui, J., Bai, S., Tian, H.} (2017). Weighted win loss 
approach for analyzing prioritized outcomes. {\it Statistics in Medicine} 
{\bf 36}, 2452--2465.
\fi 

\item {\sc Mann, H.~B. and  Whitney, D.~R.} (1947).
On a test of whether one of two random variables is stochastically larger then 
the other. {\em Annals of Mathematical Statistics} {\bf 18}, 50--60.

\item {\sc Noether, G.~E.} (1987). Sample Size Determination for Some Common 
Nonparametric Tests. {\em Journal of the American Statistical Association} 
{\bf 85}, 645--647.

\item {\sc O'Brien, R. G., and Castelloe, J. M.} (2006). Exploiting the Link 
Between the Wilcoxon-Mann-Whitney Test and a Simple Odds Statistic. {\it In: 
Proceedings of the Thirty-First Annual SAS Users Group International
Conference, Cary, NC: SAS Institute Inc.} \\
http://www2.sas.com/proceedings/sugi31/209-31.pdf .

\item
{\sc Peterson, I.} (2002). Tricky Dice Revisited. {\em Science News} 
{\bf 161}, \\ 
http://www.sciencenews.org/article/tricky-dice-revisited.

\item {\sc Pocock, S.J., Ariti, C.A., Collier, T.J., Wang, D.} (2012). The win 
ratio: a new approach to the analysis of composite endpoints in clinical trials 
based on clinical priorities. {\it European heart journal} {\bf 33}, 176--182.

\item {\sc Putter, J.} (1955). The Treatment of Ties in Some Nonparametric 
Tests. {\it The Annals of Mathematical Statistics} {\bf 26}, 368--386.

\iffalse
\item {\sc Rauch, G., Jahn-Eimermacher, A., Brannath, W., Kieser, M.} (2014). 
Opportunities and challenges of combined effect measures based on prioritized 
outcomes. {\it Statistics in Medicine} {\bf 33}, 1104--1120.

\item {\sc Sen, P.~K.} (1967). A note on asymptotically distribution-free 
confidence intervals for $Pr(X<Y )$ based on two independent samples. 
{\em Sankhya, Series A} {\bf 29}, 95--102.
\fi

\item {\sc Thangavelu, K. and Brunner, E.} (2007). Wilcoxon Mann-Whitney Test 
for Stratified Samples and Efron's Paradox Dice. {\em Journal of Statistical 
Planning and Inference} {\bf 137}, 720--737. 

\item {\sc Van Elteren, P. H.} (1960). On the Combination of Independent 
Two-Sample Tests of Wilcoxon. {\em Bulletin of the International Statistical 
Institute} {\bf 37}, 351--361.

\item {\sc Wang, D., Pocock, S.J.} (2016). A win ratio approach to comparing 
continuous non-normal outcomes in clinical trials. {\it Pharmaceutical 
Statistics} {\bf 15}, 238--245.

\item {\sc Wilcoxon, F.} (1945). Individual comparisons by ranking methods. 
{\it Biometric Bulletin} {\bf 1}, 80--83.

\item {\sc Wilcoxon, F.} (1947). Probability Tables for Individual Comparisons 
by Ranking Methods. {\it Biometrics} {\bf 3}, 119--122.

\edes

\end{document}